\documentclass[%
preprint,
showpacs, showkeys,
prd,
amsmath,amssymb,
]{revtex4-1}
\usepackage[pdftex]{graphicx}
\usepackage{bm}
\usepackage{float,color}
\usepackage{comment}
 
\newcommand{\p}[2]{\frac{\partial#1}{\partial#2}}

\newcommand{\adag}{\hat c^{\dagger}{}_{\!\!\bm{k}}}
\newcommand{\adagg}{\hat c^{\dagger}{}_{\!\!\!-\bm{k}}}
\newcommand{\anno}[1]{\textcolor{red}{#1}}
\newcommand{\bs}[1]{\boldsymbol{#1}}
\newcommand{\pa}{\partial}
\begin{document}
\title{Quantum Cramer-Rao bound for a Massless Scalar Field in de Sitter Space}
\author{Marcello Rotondo}
\email{marcello@gravity.phys.nagoya-u.ac.jp}
\author{Yasusada Nambu}%
\email{nambu@gravity.phys.nagoya-u.ac.jp}
\affiliation{%
Department of Physics, Graduate School of Science,Nagoya University,
  Chikusa, Nagoya 464-8602, Japan}
\date{October 16, 2017 ver 1.2}
%
\begin{abstract}
  How precisely can we estimate cosmological parameters by performing
  a quantum measurement on a cosmological quantum state? In quantum
  estimation theory the variance of an unbiased parameter estimator is
  bounded from below by the inverse of measurement-dependent Fisher
  information and ultimately by quantum Fisher information, which is
  the maximization of the former over all positive operator valued
  measurements. Such bound is known as the quantum Cramer-Rao
  bound. We consider the evolution of a massless scalar field
    with Bunch-Davies vacuum in a spatially flat FLRW spacetime,
  which results in a two-mode squeezed vacuum out-state for each field
  wave number mode. We obtain the expressions of the quantum Fisher
  information as well as the Fisher informations associated to
  occupation number measurement and power spectrum measurement, and
  show  the specific results of their evoluation for pure de Sitter
  expansion and de Sitter expansion followed by a radiation-dominated
  phase as examples. We will discuss these results from the point of view of the
  quantum-to-classical transition of cosmological perturbations and
  show quantitatively how this transition and the residual quantum
  correlations affect the bound on the precision.
\end{abstract}
	

\maketitle
	
\section{Introduction}
Particle creation associated to cosmological expansion is a
long-studied topic in literature \cite{Davies}. Special attention has
been paid to the case of the FLRW metrics used in cosmology
\cite{Lyth}, in particular to the maximally symmetric case of the de
Sitter spacetime for its role in modeling the early inflationary phase
of the universe and the dark-energy dominated expansion at late
times. We would like to consider the evolution of such quantum
cosmological states from the point of view of cosmological parameter
estimation.  Previous work on parameter estimation in cosmological
models has been carried out for example in \cite{Wang} for a Milne
universe described by  an expansion law with flat limit in the
asymptotic past and asymptotic future regions
\cite{Duncan}. Nevertheless, while entropic properties of scalar and
spinor fields in FLRW metric have been considered in works such as
\cite{Ball,Fuentes}, an analysis of information about the
parameters of the cosmological information is still
missing. Furthermore, while the effect of quantum-to-classical
transition of cosmological perturbations has been widely discussed
\cite{Zeh,Polarski,Campo,Kiefer}, its effect on the
information we can retrieve from them about cosmological parameters
has not yet been considered.

The main focus of the present work is to study the higher bound put on
how precisely we are allowed to estimate a generic cosmological
parameter by performing a chosen quantum measurement on the out-state
of a massless scalar Bunch-Davies vacuum in FLRW space-time. In the
specific de Sitter case, we might identify the scalar field with
perturbations of the inflaton field. In parameter estimation theory,
the lower bound for the variance of an unbiased parameter estimator
for a given measurement is provided by the Cramer-Rao bound as the
inverse of the Fisher information. In the context of quantum
estimation, quantum Fisher information is defined as the maximization
of the Fisher information over all choices of quantum measurements.
We obtain and discuss the expressions of the quantum Fisher
information and the Fisher informations for the occupation number
measurement and the power spectrum measurement, with special emphasis
on the behavior in the pure de Sitter expansion and de Sitter phase
followed by a radiation-dominated phase for the estimation of the
Hubble parameter.  Furthermore, we will provide some considerations on
how the results can be considered from the point of view of how the
quantum-to-classical transition of the inflaton field perturbations
affects the estimation precision.

The paper is structured as follows. In Sec.~\ref{1}, we quickly review
the solution of the equation of motion for a massless scalar field in
de Sitter space with Bunch-Davies vacuum and introduce the Wigner
function description of the resulting two-mode squeezed vacua.  In
Sec.~\ref{2}, we remind the reader of the definition of Fisher
information in classical and quantum estimation theory \cite{Paris}
and we provide its evaluation for the measurement of the occupation
number and the field mode amplitude of the scalar field, together with
its maximization over all quantum measurements.  In Sec.~\ref{3} we
show the specific results for the Hubble parameter estimation in the
pure de Sitter case and for the case of de Sitter expansion followed by
a radiation-dominated phase, which show significant differences.
Finally, in Sec.~\ref{4}, we address the relation of the results for
the Fisher information with the quantum-to-classical transition of
cosmological perturbations, providing final observations and
conclusions in Sec.~\ref{5}.
\section{Massless scalar field in  an expanding universe} \label{1}
Consider a massless minimally coupled scalar field $\phi$ in a
spatially flat FLRW universe of metric
\begin{equation}
	ds^2  =-dt^2+a^2\,d\bs{x}^2
        = a^2 (-d\eta^2 + d\bm{x}^2),  \label{eq:FLRW}
\end{equation}
with the scale factor $a$ and the conformal time $d\eta = a^{-1}(t)\, dt$. The
Lagrangian for this field, rescaled as $\varphi=a\,\phi$ is given by
\begin{equation}
\mathcal{L}=\int d^3x\,\frac{1}{2} \left[ \left(\varphi' - \frac{a'}{a}
  \varphi \right)^2 - (\partial_i \varphi)^2 \right] ,
\end{equation}
where $'=\partial/\partial\eta$. This Lagrangian leads to equations of
motion of the form
\begin{equation}
	\varphi'' - \left(\pa_i^2+ \frac{a''}{a} \right) \varphi= 0.
	\label{eq:eom}
\end{equation}
We introduce the Fourier mode of the field and its conjugate momentum as
\begin{align}
\varphi(\eta,{\bm{x}})= 
\int \frac{d^3k}{(2\pi)^{3/2}}\, \varphi_{\bm{k}}(\eta)\,
                     e^{i\bm{k}\cdot\bm{x}},\quad
p(\eta,{\bm{x}})  = 
\int \frac{d^3k}{(2\pi)^{3/2}}\, p_{\bm{k}}(\eta)\,
 e^{i \bm{k} \cdot \bm{x}},
\end{align}
where $p_{\bm{k}}= \varphi_{\bm{k}}'^*-(a'/a)\varphi_{\bm{k}}^*$ .
Following the usual quantization procedure, we express the field mode
and conjugate momentum in terms of the time-dependent creation and
annihilation operators $\hat{c}_{\bm{k}}(\eta)^{\dag}$ and
$\hat{c}_{\bm{k}}(\eta)$
\begin{align}
 \hat{\varphi}_{\bm{k}}  = \frac{1}{\sqrt{2k}} \left(
                        \hat{c}_{\bm{k}}
                        + \adagg \right),\quad
\hat{p}_{\bm{k}} = -i\sqrt{\frac{k}{2}}\, 
\left( \hat{c}_{\bm{k}} - \adagg \right),
	\label{eq:cre}
\end{align}
so that we can express the Hamiltonian as 
\begin{align}
\hat H = \int_{\mathbb{R}^{3+}}\!\! d^3k
\left[k\left( \hat c_{\bs{k}}{}^{\!\!\dagger} \,\hat c_{\bs{k}} 
  + \hat c_{-\bs{k}}{}^{\!\!\!\!\!\!\!\dagger}\,\, \,\hat c_{-\bm{k}} + 1 \right) + 
i\frac{a'}{a}(\adag\,\adagg-\hat c_{\bm{k}}\,\hat c_{-\bm{k}})\right] \, ,
\end{align}
where in order to treat $\bs{k}$ and $-\bs{k}$ modes as
  independent variables, integral over $\bs{k}$ is performed in half
  the Fourier space, $\bs{k}\in\mathbb{R}^{3+}$.
The Heisenberg equations of motion for the operators $\hat{c}_{\bs{k}}$ and
$\hat{c}_{\bs{k}}{}^{\!\!\dag}$ for the $\pm \bs{k}$ modes are
\begin{align}
\hat{c}_{\bm{k}}{}' &= i \left[ \hat{H} , \hat{c}_{\bm{k}}\right] 
= - i k\, \hat{c}_{\bm{k}} + \frac{a'}{a}\, \adagg\\
\adagg{}' & = i \left[ \hat{H} ,
                            {\hat{c}_{-\bs{k}}{}^{\!\!\!\!\!\!\dagger}}\,\,\right]
            = 
 i k\,\adagg + \frac{a'}{a}\, \hat{c}_{\bm{k}} .
		\label{eq:Heis_op}
\end{align}
where $k = |\bs{k}|$. In the Heisenberg formalism, the relation between
operators at a given conformal time $\eta_0$ and at a later time
$\eta>\eta_0$ is given by the Bogoliubov transformation
\begin{equation}
  \binom{\hat{c}_{\bm{k}}(\eta)}{\adagg(\eta)}
 = \binom{\alpha_k(\eta) \ \ \ \beta_k(\eta)}{\beta_k^{*}(\eta) \
   \ \ \alpha^*_k(\eta)}
 \binom{\hat{c}_{\bm{k}}(\eta_0)}{\adagg(\eta_0)}
 \, , 
	\label{eq:bog_matrix}
\end{equation}
where $\alpha_k$ and $\beta^{*}_k$ are the Bogoliubov
coefficients satisfying $|\alpha_k|^2 - |\beta_k|^2= 1$ with initial
conditions $\alpha_k(\eta_0) = 1 $ and $\beta_k(\eta_0) = 0$. The time
evolution of these coefficients is obtained straightforwardly from
\eqref{eq:Heis_op}
\begin{align}
\alpha_k{}' = - i k\,\alpha_k + \frac{a'}{a} \beta^*_k,\quad
\beta_k{}' = - i k\,\beta_k + \frac{a'}{a} \alpha_k^{*}.
		\label{eq:bogeq}
\end{align}
We define the vacuum state as the eigenstate of the annihilation
operator at time $\eta_0$
\begin{equation}
\hat{c}_{\bm{k}}(\eta_0)|\psi \rangle = 0.
\label{eq:vacuum}
\end{equation}
Introducing the Schr{\"o}dinger picture of the state at $\eta$, the
transformation \eqref{eq:bog_matrix} gives the vacuum condition
\begin{align}
  \label{eq:vacuum}
  \left( \alpha^{*}(\eta) \hat{c}_{\bm{k}} 
- \beta(\eta) \adagg \right)|\psi (\eta) \rangle = 0 ,
\end{align}
the solution of which provides the the out-state
\begin{equation}
|\psi(\eta) \rangle = \bigotimes_{k=0}^\infty \frac{1}{|\alpha_k(\eta)|}
  \sum_{n=0}^{\infty} \left(\frac{\beta_k(\eta)}{\alpha^*_k(\eta)} 
 \right)^n |n_{\bs{k}}\rangle\otimes|n_{-\bs{k}}\rangle \, .
	\label{eq:n_state}
\end{equation}
In the following, we will consider only the component for a fixed
value of $k$, as different modes do not interact with each other in
the linear order. For each value of $k$, components of
\eqref{eq:n_state} are two-mode ($\pm\bs{k}$) squeezed vacuum states,
with squeezing parameters $r_k,\theta_k, \phi_k$ related to the
Bogoliubov coefficients by
\begin{align}
  \alpha_k=e^{-i\theta_k}\cosh
  r_k,\quad\beta_k=e^{i(\theta_k+2\phi_k)}\sinh r_k \, ,
\end{align}
$\theta_k$ being a free phase. For simplicity of notation, in the
following we will omit the index $k$ in the parameters. 

 The squeezing
magnitude $r$ and phase $2\phi$ follow the evolution equations
\begin{equation}
    r'=\frac{a'}{a}\cos 2\phi,\quad\phi'=-k-\frac{a'}{a}\coth
    2r\sin2\phi,\quad
    \theta'=k+\frac{a'}{a}\tanh r\sin 2\phi \, .
\end{equation}
In the basis that diagonalizes
$\{\hat{\varphi}_{\bs{k}},\hat{\varphi}_{-\bs{k}}\}$, the vacuum
condition \eqref{eq:vacuum} provides the Gaussian wave function
\begin{equation}
  \psi(\varphi_{\bs{k}},\varphi_{-\bs{k}})=\langle\varphi_{\bs{k}},\varphi_{-\bs{k}}|\psi(\eta)\rangle_{\bs{k}}\propto\exp\left[-\gamma\,\varphi_{\bs{k}}\varphi_{-\bs{k}}\right], 
\end{equation}
where $\gamma=\beta/\alpha*$. In order to deal with real coordinates,
in the following we will introduce Hermitian field mode variables
$\hat q_{\bm{k}}$ and $\hat \pi_{\bm{k}}$ as
\begin{align}
  &\hat
    q_{\bs{k}}=\frac{1}{2}\left[\hat\varphi_{\bs{k}}+\hat\varphi_{-\bs{k}}+
    \frac{i}{k}\,(\hat p_{\bs{k}}-\hat p_{-\bs{k}})\right],\\
  &\hat\pi_{\bs{k}}=\frac{1}{2}\left[\hat p_{\bs{k}}+\hat
    p_{-\bs{k}}-ik\,
    (\hat\varphi_{\bs{k}}-\hat\varphi_{-\bs{k}})\right] \, .
\end{align}
and analogously for the variables for $-\bs{k}$.  In the basis that
diagonalizes $\{\hat q_{\bs{k}},\hat q_{-\bs{k}}\}$, the wave function
reads
\begin{align}
	&\psi(q_{\bs{k}},q_{-\bs{k}})=\langle
          q_{\bs{k}},q_{-\bs{k}} | \psi(\eta) \rangle_k
\propto \exp\! \Big[ - \frac{k}{4} \left( \frac{1-\gamma}{1+\gamma}
  \right) (q_{\bs{k}} + q_{-\bs{k}})^2  
	 - \frac{k}{4} \left( \frac{1+\gamma}{1-\gamma} \right)
  (q_{\bs{k}} - q_{-\bs{k}})^2 \Big],
\label{eq:Wfunc}
\end{align}
where $\gamma=\beta/\alpha^*$.
It is convenient to consider the Wigner function associated to the
wave function of the pure state $\psi(q_{\bs{k}},q_{-\bs{k}})$ for
which explicit calculation gives the Gaussian distribution
  \begin{align}
    W(q_{\bs{k}},\pi_{\bs{k}},q_{-\bs{k}},\pi_{\bs{k}})&\equiv
    \frac{1}{\pi^2}\int_{-\infty}^{\infty} dx\,dy\, e^{-2i(\pi_{\bs{k}}\,x+\pi_{-\bs{k}}\,y)}
    \,\psi(q_{\bs{k}}+x,q_{-\bs{k}}+y)\,\psi^*(q_{\bs{k}}-x,q_{-\bs{k}}-y)
                                                                   \notag
    \\
    &=\frac{1}{\pi^2}\exp\left(-\frac{1}{2}\bs{\xi}^TV^{-1}\bs{\xi}\right), 
     \quad
      \bs{\xi}^T\equiv(q_{\bs{k}},\pi_{\bs{k}},q_{-\bs{k}},\pi_{-\bs{k}})
      \label{eq:wigner}
  \end{align}
where the covariance matrix is
$V_{ij}=\langle\xi_i\xi_j+\xi_j\xi_i\rangle/2$, which specifies
completely the state and can be expressed in the block form as
\begin{equation}
  V=  \frac{1}{2}\cosh(2r) \begin{pmatrix}
   \mathbb{I} & \tanh(2r) \left( \cos(2\phi) \sigma_3 + \sin(2\phi)
     \sigma_1 \right) \\ 
  \tanh(2r)\left( \cos(2\phi) \sigma_3 + \sin(2\phi) \sigma_1 \right)
  & \mathbb{I} 
  \end{pmatrix},
    \label{eq:covariance}
\end{equation}
$\sigma_1$ and $\sigma_3$ being Pauli matrices and $\mathbb{I}$
the $2\times2$ identity matrix.
\section{Fisher Information of the scalar field} \label{2}
In order to discuss the bound on the estimation precision\anno{,} we
consider the Fisher information.  In the classical case, let
$P(x;\theta)$ be the probability distribution to obtain a certain
measurement result $x$, parametrized by a single scalar $\theta$. The
Cramer-Rao bound provides a lower bound for the variance of the
estimator $\hat{\theta}_\text{est}(\{x\})$ of the parameter $\theta$
through the single measurement of a variable $x$
\begin{equation}
\langle (\Delta{\hat\theta})^2\rangle \geq \frac{1}{F(\theta)}  ,
\end{equation}
where $\Delta{\hat\theta}=\hat{\theta}_\text{est}(\{x\})-\theta$ and
$F(\theta)$ is the Fisher information defined by
\begin{equation}
	F(\theta) = \int_X dx\,P(x;\theta) \left( \p{\log P(x;\theta)
          }{\theta}\right)^2 \, . 
\end{equation}
For simplicity we consider a single scalar parameter $\theta$, but the
extension to the multiparameter case is straightforward.

In the context of quantum estimation theory \cite{Paris}, the Fisher
information can be expressed in terms of a positive operator-valued
measurement $\{\hat{\Pi}_x\}$. For a state $\hat\rho$ parametrized by the
quantity $\theta$, the result of the measurement has probability
distribution $P(x;\theta)=\mathrm{Tr}(\hat{\Pi}_x\,\hat\rho)$ and the
Fisher information is
\begin{equation}
F_{\hat{\Pi}_x}(\theta) = \int dx
   \frac{\mathrm{Re}\left(\mathrm{Tr}
    [\,\hat\rho\,\hat{\Pi}_x\hat{L}_{\theta}]\right)^2}{\mathrm{Tr}
[\,\hat\rho\,\hat{\Pi}_x]},
\label{eq:CF_quantum}
\end{equation}
where one makes use of the symmetric logarithmic derivative (SLD)
$\hat L_\theta$, a self-adjoint operator implicitly defined by
\begin{equation}
\p{\hat\rho}{\theta} = \frac{1}{2} \left( \hat{L}_\theta \hat\rho
 + \hat\rho\hat{L}_\theta  \right)\, . 
\end{equation}
The Fisher information can be shown to be maximized over all positive
operator-valued measurements $\{\hat\Pi_x\}$ by the quantum Fisher
information
\begin{equation}
 F_Q(\theta) = \mathrm{Tr}[\,\hat\rho\,{\hat{L}_\theta}^2] = 
\mathrm{Tr}[\partial_\theta \hat\rho\, {\hat{L}_\theta}] ,
 \label{eq:FQ}
\end{equation}
which provides the ultimate quantum Cramer-Rao bound
\begin{equation}
 \langle (\Delta\hat{\theta})^2\rangle\geq \frac{1}{F_{\hat\Pi_x}(\theta)}
 \geq \frac{1}{F_Q(\theta)}
 \label{eq:qcr_bound}
\end{equation}
with the equality being satisfied when the measurement $\{\hat{\Pi}_x\}$
projects over the eigenstates of the SLD operator.  For a pure state
$\hat\rho=|\psi\rangle\langle\psi|$, $\hat\rho=\hat\rho\,\hat\rho$ and
\begin{equation}
  \frac{\partial\hat\rho}{\partial\theta}=
\frac{\partial\hat\rho}{\partial\theta}\,\hat\rho+\hat\rho\,
\frac{\partial\hat\rho}{\partial\theta},
\end{equation}
hence the SLD operator takes the simple form
\begin{equation}
 \hat{L}_\theta = 2\, \p{\hat{\rho}}{\theta} \, .
 \label{eq:L}
\end{equation}
Therefore, for the pure state, the quantum Fisher information can be
expressed as
\begin{equation}
  F_\text{Q}=4\Bigl[\langle\pa_\theta\psi|\pa_\theta\psi\rangle-\langle\psi|
\pa_\theta\psi\rangle\langle\pa_\theta\psi|\psi\rangle\Bigr] \, .
 \label{eq:QFI}
\end{equation}
For our purposes, we can restrict ourselves to consider a pure
Gaussian state (i.e. a quantum state with Gaussian Wigner function)
whose wave function and Wigner function are given by \eqref{eq:Wfunc}
and \eqref{eq:wigner}, respectively (for a reference on quantum
estimation theory using single-mode Gaussian state see
e.g. \cite{Pinel}) . By taking derivative of the Wigner
function $W$ with respect to a parameter
$\theta$,
\begin{align}
  &\pa_\theta W(q_{\bs{k}},\pi_{\bs{k}},q_{-\bs{k}},\pi_{-\bs{k}})
    \notag \\
 &=\int_{-\infty}^{\infty}
  dx\, dy\,e^{2i(\pi_{\bs{k}}\,x+\pi_{-\bs{k}}\,y)}\,\pa_\theta\left[
    \psi(q_{\bs{k}}-x,q_{-\bs{k}}-y)\,\psi^*(q_{\bs{k}}+x,q_{-\bs{k}}+y)\right],
\end{align}
it is possible to show that
\begin{align}
  \int dq_{\bs{k}}\,dq_{-\bs{k}}\,d\pi_{\bs{k}}\,d\pi_{-\bs{k}}
\left(\pa_\theta W\right)^2&=
         \frac{1}{2\pi^2}\Bigl[\langle\pa_\theta\psi|\pa_\theta\psi\rangle
           +\langle\psi|\pa_\theta\psi\rangle^2 \Bigr],
\end{align}
and the quantum Fisher information can be expressed in terms of the
covariance matrix as
\begin{equation}
  F_Q=8\pi^2\int dq_{\bs{k}}\,dq_{-\bs{k}}\,d\pi_{\bs{k}}\,d\pi_{-\bs{k}}
\,(\pa_\theta W)^2=
  \frac{1}{4}\,\mathrm{Tr}\Bigl[\left((\pa_\theta V)V^{-1}\right)^2\Bigr].
\end{equation}
For the covariance matrix \eqref{eq:covariance}, we obtain
\begin{equation}
  F_Q=
  4(\alpha^*\pa_\theta\beta-\beta\pa_{\theta}\alpha^*)(\alpha\pa_\theta\beta^*
  -\beta^*\pa_\theta\alpha)=4\Bigl[(\pa_\theta r)^2+(\sinh 2r)^2 (\pa_\theta\phi)^2\Bigr].
  \label{eq:f_q}
\end{equation}

Let us now consider specific measurements of the scalar field. A first
natural choice could be to evaluate the Fisher information for  the
projective measurement of the occupation number,
\begin{equation}
\hat{\Pi}_n =|n_{\bm{k}}\rangle\langle
  n_{\bm{k}}|\otimes |n_{-\bm{k}}\rangle\langle n_{-\bm{k}}| .
\end{equation}
The state after the measurement is
\begin{equation}
\hat\rho_{\text{cl}}=\sum_{n=0}^{\infty}P(n)|n_{\bs{k}}\rangle{}\langle
n_{\bs{k}}|\otimes|n_{-\bs{k}}\rangle{}\langle n_{-\bs{k}}| 
\label{eq:clstate}
\end{equation}
with measurement outcome probability
\begin{equation}
P(n)=\mathrm{Tr}(\hat\Pi_n\,\hat\rho) = \frac{1}{\vert \alpha \vert^2}
\left|\frac{\beta}{\alpha} \right|^{2n} \, .
\label{eq:fin}
\end{equation}
We found that the Fisher information associated to \eqref{eq:fin} is simply
\begin{align}
	F_{n}(\theta) = 4 \left(\pa_\theta \ln |\alpha| \right) \left(\pa_\theta \ln |\beta|\right) = 4(\pa_\theta r)^2  \, . 
	\label{eq:f_n}
\end{align}
This shows that the number measurement is allowed to access the part
of quantum Fisher information associated to the parameter-dependency
of the squeezing magnitude, but not that associated to the
parameter-dependency of the squeezing phase.
	%
In other words, the number measurement allows the highest precision
for the estimation of a cosmological parameter as long as the
squeezing phase of the state does not depend on that parameter, which
generally does.

To have a better insight into the effect of decoherence on the allowed
estimation precision, which we will discuss in Sec.~V, let
us consider also the measurement of the power spectrum of the field
(i.e. the modulus squared of the amplitude of its modes). For this
purpose, we introduce new variables as
\begin{equation}
  2X_1=q_{\bs{k}}+q_{-\bs{k}},\quad 2X_2=q_{\bs{k}}-q_{-\bs{k}},\quad
  2Y_1=\pi_{\bs{k}}+\pi_{-\bs{k}},\quad 2Y_2=\pi_{\bs{k}}-\pi_{-\bs{k}}\, .
\end{equation}
The original field mode $\varphi_{\bs{k}}$ is expressed as
\begin{equation}
  \varphi_{\bs{k}}=X_1+\frac{i}{k}Y_2,\quad
  \varphi_{\bs{k}}\,\varphi_{\bs{k}}{}^*=
  (X_1)^2+\frac{1}{k^2}(Y_2)^2 \, .
\end{equation}
From the Wigner function \eqref{eq:wigner}, by integrating over $X_2,
Y_1$, the marginal probability distribution for variables $X_1, Y_2$ is
obtained
\begin{equation}
  P(X_1,Y_2)=\frac{2}{\pi(\cosh 2r+\cos 2\phi\sinh 2r)}\exp\left[
    -\frac{2(k^2X_1^2+Y_2^2)}{k(\cosh 2r+\cos 2\phi\sinh
      2r)}\right] \, .
  \label{eq:Pcl}
\end{equation}
As the operators associated to $X_1,Y_2$ commute, it is possible to
measure these variables simultaneously by homodyne detection for the
quadratures $\hat{X}_1,\hat{Y}_2$ (notice that choice of variables
$X_2,Y_1$ corresponds to the measurement of conjugate momentum
amplitude). The Fisher information for the probability distribution
\eqref{eq:Pcl} is
\begin{equation}
F_{\varphi}(\theta)=4\left(\pa_\theta \ln |\alpha+\beta^*|\right)^2=4\left(\frac{\tanh 2r+\cos 2\phi}{1+\cos
	2\phi\tanh 2r}\pa_\theta r-\frac{\sin 2\phi\tanh 2r}
{1+\cos 2\phi\tanh 2r}\pa_\theta\phi\right)^2
\label{eq:f_f}
\, ,
\end{equation}
where the contributions from parameter-dependency of the squeeze
magnitude and squeeze phase are now mixed.

Notice that the Fisher information for the power spectrum measurement
is related to the strength of correlations between the $+\bs{k}$ and
$-\bs{k}$ modes, which are given by the expectation value for the
field mode amplitude
\begin{equation}
   \langle\varphi_{\bs{k}}\,\varphi_{-\bs{k}}\rangle =
   \langle\varphi_{\bs{k}}\, \varphi_{\bs{k}}{}^*\rangle
   =\int_{-\infty}^{\infty} dX_1 dY_2
  \left(X_1^2+\frac{Y_2^2}{k^2}\right)P(X_1,Y_2)=\frac{1}{2k}
  (\cosh 2r+\cos 2\phi\sinh 2r)\, 
\label{eq:qcol}
\end{equation}
since $\varphi_{\bs{k}}^* = \varphi_{-\bs{k}}$. By writing explicitly
the wave function in the $\varphi_{\bs{k}},\varphi_{\bs{k}}{}^*$
representation, it is possible to show that these correlations can be
expressed in terms of the Bogoliubov coefficients simply as
\begin{equation}
\langle\varphi_{\bs{k}}\,\varphi_{-\bs{k}}\rangle 
= \frac{1}{2k} \left| \alpha_k + \beta^*_k \right|^2
\end{equation}
and therefore from \eqref{eq:f_f} one has 
\begin{equation}
F_{\varphi} (\theta)
 = \left( \pa_\theta \ln \left(
     \langle\varphi_{\bs{k}}\,\varphi_{-\bs{k}}\rangle\right) \right)^2
 \, . 
\end{equation}
The correlations \eqref{eq:qcol} may be compared to the ones
associated to the projection of the state on the number basis
\begin{equation}
\langle\varphi_{\bs{k}}\,\varphi_{-\bs{k}}\rangle_{\text{cl}}
=\frac{1}{2k}\cosh(2r) \, .
	\label{eq:clcol}
\end{equation}
In the limit of large squeezing, these correlations are related as
\begin{equation}
	\langle \varphi_{\bs{k}} \varphi_{-\bs{k}}\rangle
  \simeq (1+\cos2\phi) \langle \varphi_{\bs{k}}
  \varphi_{-\bs{k}}\rangle_{\text{cl}} \, . 
		\label{eq:lim_col}
\end{equation}
The correlations \eqref{eq:qcol} and \eqref{eq:clcol} are related to
the Shannon entropy of the respective measurement distributions as
follows. The probability density function of the power spectrum of the
field is
\begin{equation}
P(|\varphi_{\bs{k}}|^2) = \frac{1}{2\langle \varphi_{\bs{k}}
  \varphi_{-\bs{k}}\rangle}\exp\left[ - \frac{1}{2}
  \frac{|\varphi_{\bs{k}}|^2}{\langle \varphi_{\bs{k}}
    \varphi_{-\bs{k}}\rangle}\right] \, ,
\end{equation}
which has the following Shannon entropy  
\begin{equation}
S_\varphi =
  - \int_0^\infty d \left( |\varphi_{\bs{k}}|^2 \right)
  P(|\varphi_{\bs{k}}|^2) \ln \left(P(|\varphi_{\bs{k}}|^2) \right) =
  1 + \ln 2 +  \ln\langle \hat \varphi_{\bs{k}} \hat
  \varphi_{-\bs{k}}\rangle \ \, .
\label{eq:S_f}
\end{equation}
This provides us with the simple relation between the Fisher
  information and the Shannon entropy for field mode amplitude
\begin{equation}
	F_{\hat{\varphi}}(\theta)
 = \left( \pa_\theta S_{\varphi} \right)^2 \, .
	\label{eq:FS_f}
\end{equation}
The entropy for the number state projection  is
\begin{align}
	S_\text{cl }&= - \sum_{n=0}^\infty P(n)\log P(n) 
        =|\alpha|^2\log|\alpha|^2-|\beta|^2\log|\beta|^2  \notag \\
		 &= \cosh(r)^2\ln \cosh(r)^2-\sinh(r)^2\ln \sinh(r)^2,
\end{align}
which coincides with the entropy of entanglement between $+\bs{k}$ and
$-\bs{k}$ modes.  In the large squeezing limit then we have
\begin{equation}
S_\text{cl} \simeq  S_\varphi \simeq 2r \, .
\label{eq:s_bound}
\end{equation}
As we will explain in Sec.~V, the value $2r$
corresponds to the maximal attainable value of the entropy via coarse
graining of the state with complete randomization.  Notice that the
result \eqref{eq:FS_f} is exact and relates directly the Fisher
information for the power spectrum measurement to the field entropy.

\section{Fisher information in de Sitter universe} \label{3} 
\begin{figure}[H]
    \centering
    \includegraphics[width=0.49\linewidth,clip]{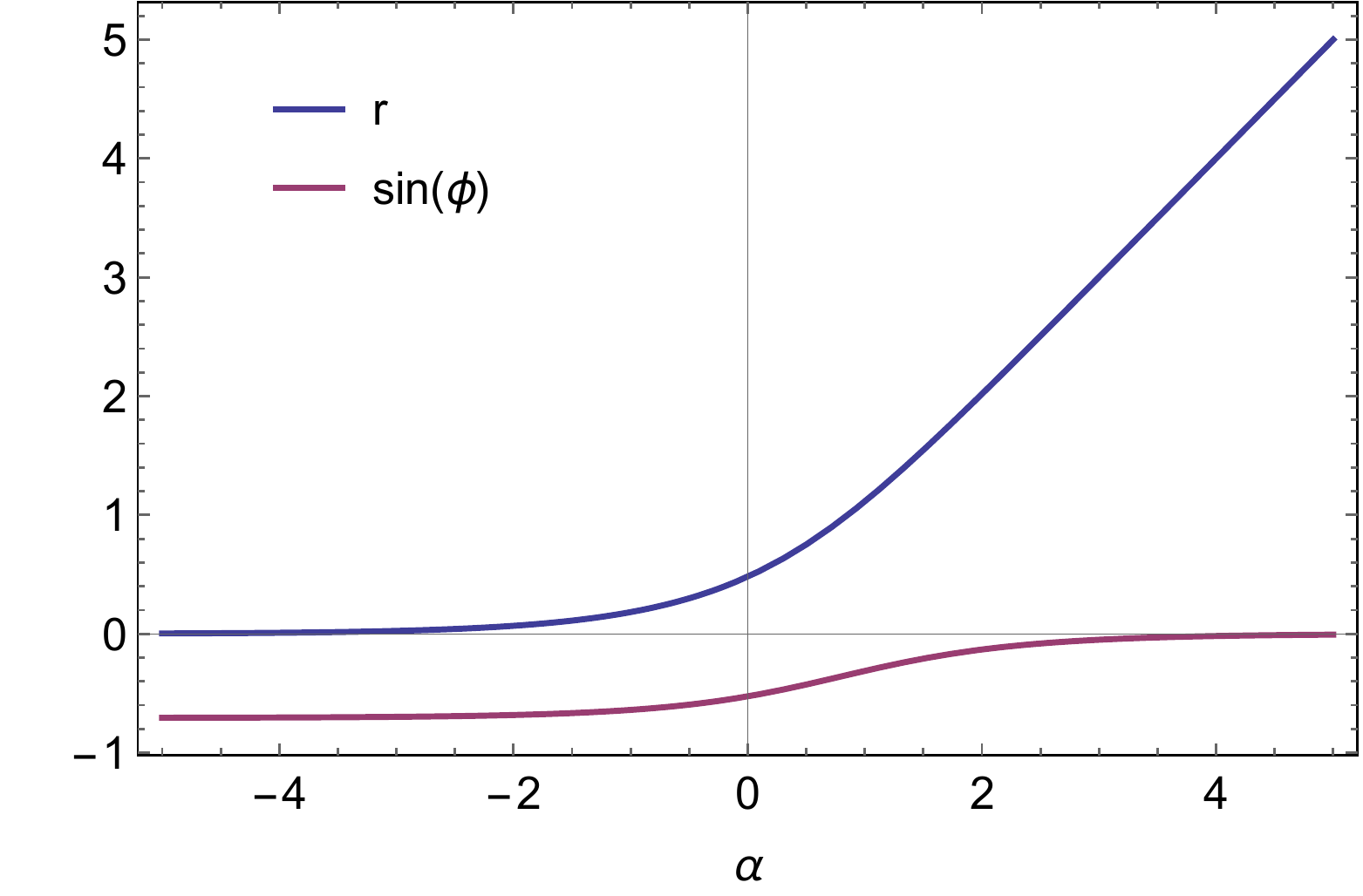}
    \includegraphics[width=0.50\linewidth]{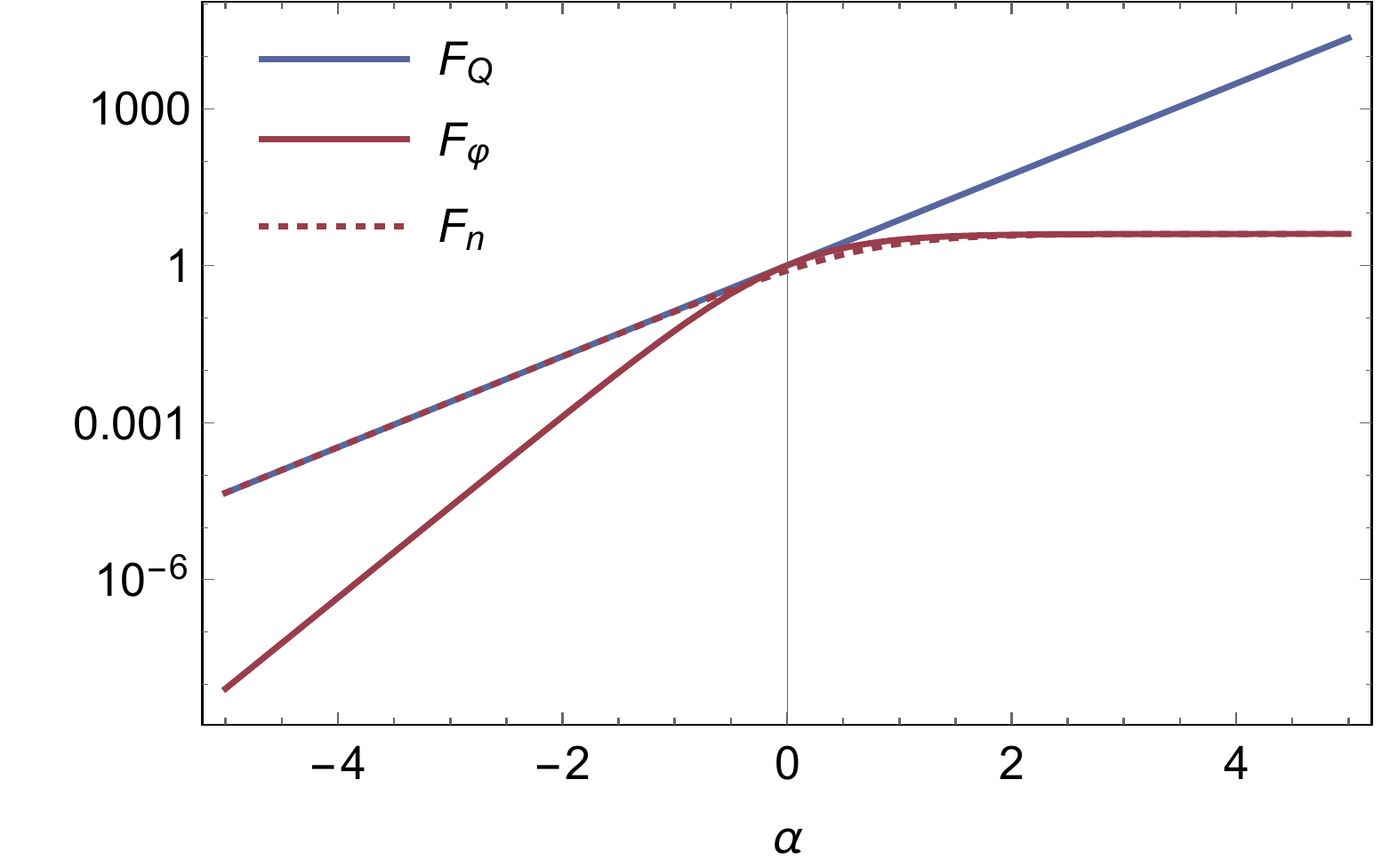}
    \caption{Left: Evolution of squeezing parameter and squeezing
      phase for $k=1$.
      Right: Evolution of Fisher informations divided by $G^2$
      for $k=1$.}
   \label{fig:dssq}
\end{figure}

As an example of the results of the previous section, let us consider
a pure de Sitter space-time with conformal factor 
$a(\eta)=-1/(H\eta), -\infty<\eta<0$. For the Bunch-Davies vacuum, the
Bogoliubov coefficients are
\begin{equation}
	\alpha_k = \left(1  - \frac{i}{2 k \eta}  \right) e^{-ik\eta},\quad
	\beta_k = \frac{i}{2k\eta} e^{ik\eta} \, .
	\label{eq:bog}  
\end{equation}
The squeezing parameters are
\begin{equation}
    r_k=-\sinh^{-1}\left(\frac{1}{2k\eta}\right),\quad\phi_k=-\frac{\pi}{4}
-\frac{1}{2}\tan^{-1}\left(\frac{1}{2k\eta}\right)\, .
\end{equation}
Fisher informations are given by
\begin{equation}
    F_Q=\frac{e^{2\alpha}H^2}{k^2}G(\theta)^2,\quad
    F_{\varphi}=4\left(\frac{e^{2\alpha}H^2}{e^{2\alpha}H^2+k^2}\right)^2
    G(\theta)^2,\quad
    F_{n}=\frac{4e^{2\alpha}H^2}{e^{2\alpha}H^2+4k^2}G(\theta)^2,
\end{equation}
where $\alpha=\log
a$ is the $e$-foldings of cosmic expansion and $G(\theta) =
\pa\alpha/\pa\theta$ represents the parameter dependence of the
$e$-foldings.
Fig.~\ref{fig:dssq}(left panel) shows the evolution of the squeezing
parameter and squeezing phase. When the physical wavelength is smaller
than the Hubble horizon $a/k<H^{-1}$
(i.e. for modes inside the Hubble horizon), the squeezing parameter
and phase are constant. After the horizon exit at $\alpha=0$,
we have $a/k>H^{-1}$ and $r$ grows while $\phi$ approaches zero.
\noindent
In the super-horizon scale $k/a\ll
H$, these Fisher informations behave as
\begin{equation}
 F_{\varphi} \approx F_{n}\ll F_Q.
\end{equation}
On the other side, in the small scale $k/a \gg H$, we have
\begin{align}
 F_{n} \approx 
\left(\frac{H}{k/a}\right)^{2}G(\theta)^2=F_Q,\quad
 F_{\varphi}\approx 4\left(\frac{H}{k/a}\right)^4G(\theta)^2,
\end{align}
thus, we obtain the relation
  \begin{equation}
    F_{\varphi}\ll F_{n}\approx F_Q.
  \end{equation}
  As the state is squeezed more and more in the super horizon scale,
  most amount of the quantum Fisher information becomes inaccessible
  to either of the measurements. As we discussed before, the
  difference between $F_Q$
  and $F_{n}$
  is due to the parameter dependency of the squeezing phase
  $\phi$.
  As one may notice by looking at the covariance matrix
  \eqref{eq:covariance}, obtaining such information requires a
  measurement that properly accesses the correlations between the
  $+\bs{k}$
  and $-\bs{k}$
  modes, represented by the off-diagonal blocks of the matrix. Notice
  that for any given $k$-mode,
  the squeezing phase $\phi$
  is related to the quantum phases of the two modes of the squeezed
  vacuum in the number representation by $\langle
  \hat{\phi}_{+} + \hat{\phi}_{-} \rangle = 2\phi
  $, where
  $\hat{\phi}_{\pm}$
  is the operator appearing at exponent in the Pegg-Barnett phase
  operator, acting on the $\pm\bs{k}$ modes \cite{Gantsog}.

\begin{figure}[H]
    \centering
    \includegraphics[width=0.47\linewidth,clip]{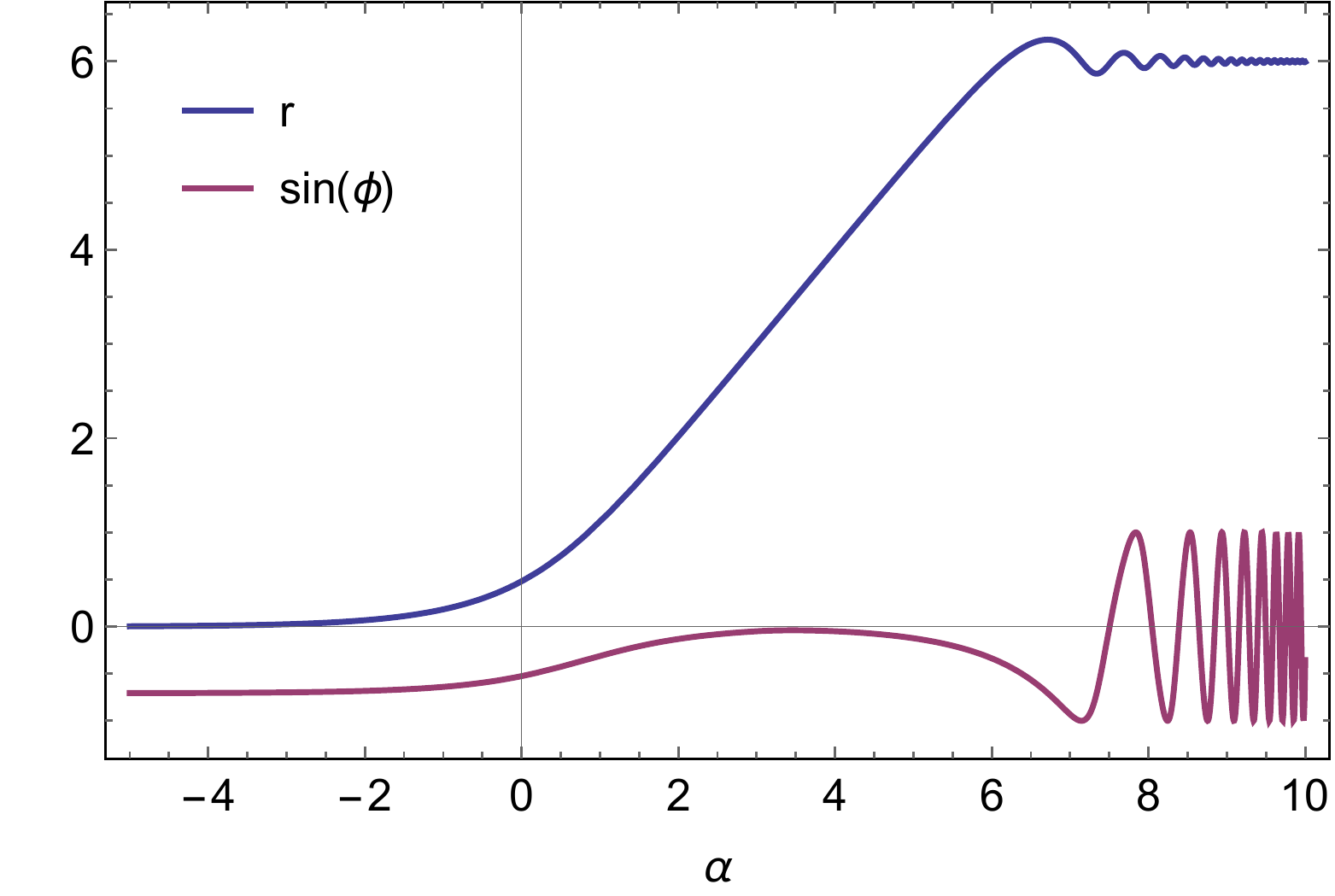}
 \includegraphics[width=0.52\linewidth]{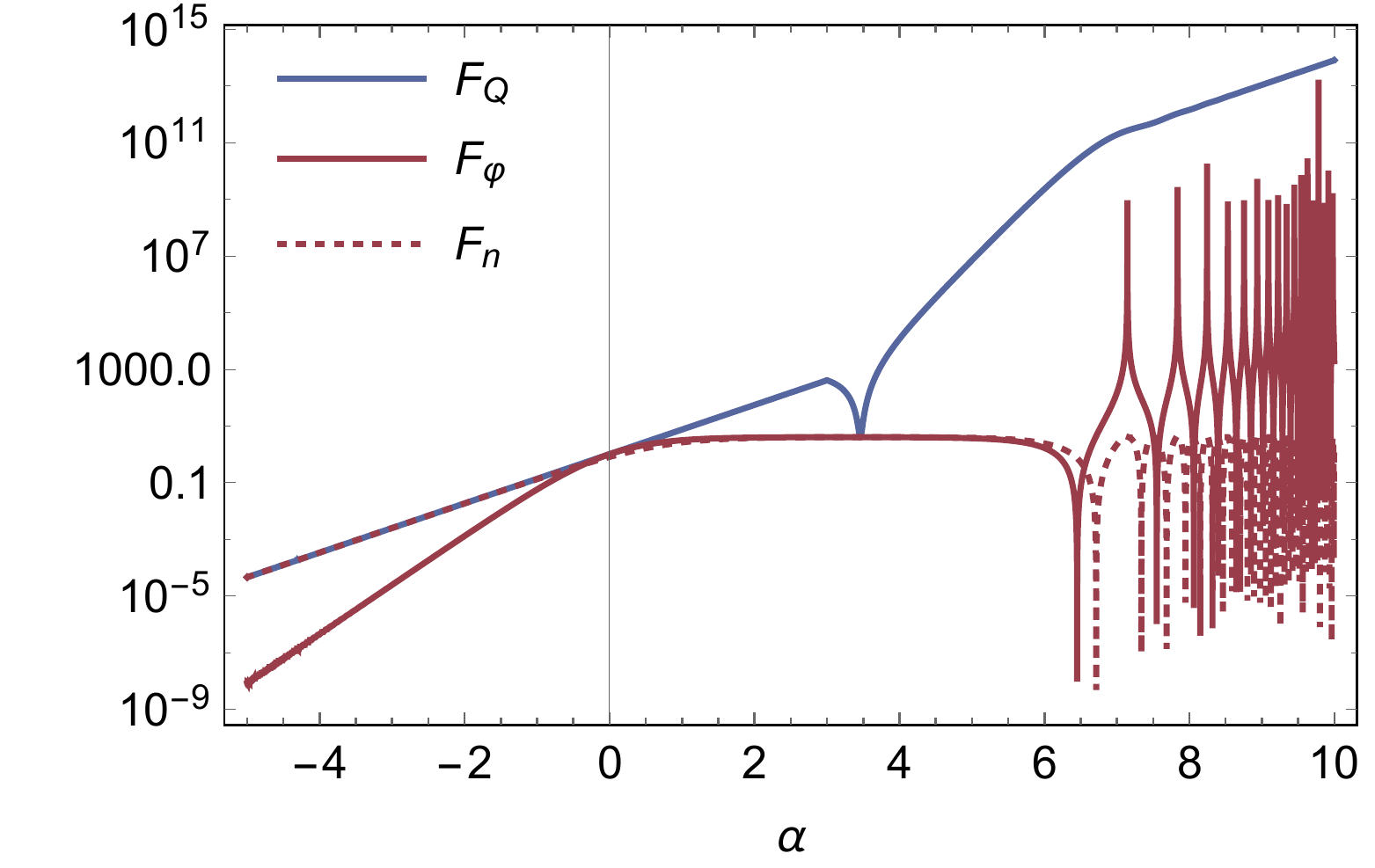} \centering
    \caption{Left: Evolution of squeezing parameters for $k=1$. Right:
      Evolution of Fisher informations divided by $G^2$ for $k=1$.}
   \label{fig:dsrsq}
\end{figure}

Now let us consider a cosmological model with transition
from de Sitter expansion to decelerated expansion (radiation dominated
phase). The scale factor in this case is
\begin{equation}
    a(\eta)=
    \begin{cases}
        -1/(H\eta)&\quad\text{for}\quad \eta\le\eta_1<0\\
        (\eta-2\eta_1)/(H\eta_1^2)&\quad\text{for}\quad \eta>\eta_1
    \end{cases} \, .
\end{equation}
Fig.~\ref{fig:dsrsq} shows evolution of squeezing parameters and
Fisher informations for the $k=1$ mode. The transition from de Sitter
expansion to decelerated expansion is at $\eta_1=-1/e^3$ which
corresponds to $\alpha_1=3$.
\noindent 
During the de Sitter phase, the physical wavelength of this mode
exceeds the Hubble horizon length. After $\alpha\approx 6$, the
wavelength becomes again smaller than the Hubble horizon length and,
after this time, the squeezing magnitude and phase start to oscillate,
as already observed in \cite{Lohmar}. After the reentry of the mode in
the Hubble horizon, Fisher informations $F_{\varphi}$ and
$F_{n}$ also oscillate. In particular, the Fisher information
$F_{\varphi}$ reaches the same values as $F_Q$ for some
values of $\alpha$. This means that there is the possibility that the
estimation through measurement of the power spectrum can attain the
quantum Cramer-Rao bound.  The reason for this is that in the
radiation dominated era, a contour of the Wigner function becomes an
elongated ellipse with fixed axises (since $r$ approaches a constant
value) and rotates in the phase space in such a way that at some point
it becomes completely accessible to the measurement of quadratures
$\hat{X}_1,\hat{Y}_2$. The equality $F_\text{Q}=F_{{\hat{\varphi}}}$
holds for
\begin{equation}
 \phi \approx -\frac{1}{2}\tan^{-1}\left(\frac{1}{\cosh
     2r}\right)+n\pi,\quad n=0,\pm 1,\pm2,\cdots \, .
 \label{eq:sat}
\end{equation}
Thus if we consider the homodyne measurement of the field quadrature,
we have a chance to attain the quantum Cramer-Rao bound periodically
in time.  On the other hand, the amplitude of $F_{n}$ keeps
the same maximal value when the mode is outside of the Hubble horizon
and this value is far smaller than the value of $F_Q$ in the
radiation dominated era.

\section{Quantum-to-classical transition of cosmological perturbations} \label{4}

In the previous sections, we have considered the amount of quantum
Fisher information of a cosmological parameter that is accessible by
performing two different projective measurements on the pure
cosmological state. In such case, the fact that not all of the quantum
Fisher information for the parameter can be accessed is ascribed to
the fact that the choice of measurement is non optimal, in the sense
that it does not satisfy the equality in the quantum Cramer-Rao bound
\eqref{eq:qcr_bound}.
Now we would like to adopt a different point of view and see the
projection over the chosen set of states as representing the
interaction of the cosmological state with an external environment
rather than with an observer. From this point of view, the fact that
quantum Fisher information becomes partially inaccessible is ascribed
to the entanglement of the initially pure state with the degrees of
freedom of the environment, which are inaccessible to the observer. As
a consequence, to such an observer the state will appear not pure but
mixed.

The motivation behind this slight change of perspective is that the
field mode amplitude states that we considered can be taken
approximatively as the pointer basis of the evolution of cosmological
perturbations, i.e. the basis ``naturally chosen'' by the interaction
with the environment leading from quantum perturbations to classical
ones \cite{Lohmar}. This environment-induced interpretation of the
information loss requires the interaction to be strong enough as to be
analogous to a instantaneous projective measurement. Such interaction
of the cosmological system with the environment is a realistic
consideration, in particular when one takes into account the frailty
of squeezing \cite{Polarski}. The interaction with the environment can
be generically modeled by attaching a dumping factor to the initial
pure state $\rho_0$
\begin{equation}
  \rho_\xi (\varphi_{\bs{k}},\varphi'{}_{\!\!\bs{k}}) = \rho_0
  (\varphi_{\bs{k}},\varphi'{}_{\!\!\bs{k}}) \exp\left( -\frac{\xi}{2}
    \, |\varphi_{\bs{k}}-\varphi'{}_{\!\!\bs{k}}|^2\right) \, , 
		\label{eq:xi}
\end{equation}
$\xi$ being a parameter that encodes the details of the
interaction. The decoherence from the initial pure state to the final
mixed state can be quantified by the von Neumann entropy
$S^{(\text{vN})}(\rho_\xi)$ of the total state \eqref{eq:xi}. In the limit
$\xi \langle \hat\varphi_{\bs{k}}\, \hat\varphi_{-\bs{k}}\rangle \gg 1$,
\cite{Lohmar} found
\begin{equation}
  S^{(\text{vN})}(\rho_\xi)  \simeq  1 - \ln 2 + \frac{1}{2}\ln \xi \langle
  \hat\varphi_{\bs{k}} \hat\varphi_{-\bs{k}}\rangle    \, , 
\end{equation}
when expressed in the present notation. By substituting
\eqref{eq:S_f}, taking the limit for large squeezing, we have the
following relation
\begin{equation}
	S^{(\text{vN})}(\rho_\xi) \leq S_\text{max}/2\simeq S_{\varphi}/2 \simeq
        S_{\text{cl}}/2  \, ,
	\label{eq:S_f_xi}
\end{equation}
where $S_{\text{max}}$ represents the maximum value of the entropy for
this bipartite system. In terms of the Wigner function, this value is
obtained by smearing the Wigner ellipse to become a circle. As
discussed in \cite{Lohmar}, such a bound shows that the reduction over
the field amplitude pointer states does not lead to complete quantum
decoherence but retains some amount of ``quantumness''. 

From the point of view of decoherence, as a result of the
quantum-to-classical transition of cosmological perturbations, the
quantum Fisher information for the local system is reduced from the
initial $F_Q(\theta)$ to a lower value
$F_Q(\theta)' = F_{\varphi}(\theta)$ after the transition, power
spectrum becoming the optimal measurement in terms of parameter
estimation. Corresponding to this decrease in quantum Fisher
information, there is an increase in the von Neumann entropy from its
minimal value of zero for the initial pure state to its maximal value
of $S_{\text{max}}/2$.  Notwithstanding this cosmological decoherence,
the interaction with the environment does not lead to total loss of
quantum correlations, whose contribution to the final quantum Fisher
information $F_Q(\theta)'$ can be determined as
$F_{\varphi}(\theta) - F_n(\theta)$. In the radiation dominated case,
such contributions periodically allow to access all the Fisher
information of the cosmological parameter that presented before the
quantum-to-classical transition through a measurement of power
spectrum, while this information is lost in the limit of pure de
Sitter space. Notice that, consistently, in the large squeezing limit,
the entropy shows log-periodic fluctuations due to the factor
$\cos\phi$ in \eqref{eq:lim_col} and such fluctuations disappear in
the large squeezing limit of the de Sitter case, where $\phi \to 0$
and
$\langle \varphi_{\bs{k}} \varphi_{-\bs{k}}\rangle \simeq 2 \langle
\varphi_{\bs{k}} \varphi_{-\bs{k}}\rangle_{\text{cl}} $.

One may observe then that the fact that the mechanism of cosmological
decoherence is not perfectly efficient turns out to be extremely
important from the point of view of observation, not only for
generating observable features such as the acoustic peaks in the
anisotropy spectrum of the CMB (as observed in \cite{Kiefer}) but also
for being determinant in how well we can estimate cosmological
parameters from our observations by accessing quantum correlations by
a proper choice of the measurement.
\section{Conclusions} \label{5}

In conclusion, constraining ourselves to the case of the massless
Bunch-Davies scalar vacuum in a FLRW metric parametrized by some
scalar cosmological parameter, we found the explicit expressions for
the quantum Fisher information and the Fisher informations associated
to the measurement of the power spectrum and occupation number, both
in terms of the Bugoliubov coefficients of the cosmological unitary
transformation and in terms of the squeezing magnitude and phase of
the two-mode squeezed out-vacuum state. While quantum Fisher
information cannot be accessed completely by the two chosen
measurements in the pure de Sitter case, it can be accessed
periodically by the power spectrum measurement when we introduce a
radiation-dominated phase, such periodicity being related to the
rotation of the Wigner function in the quantum phase space. In the
context of the quantum-to-classical transition of cosmological
perturbations, we have shown quantitatively how the transition affects
the value of the quantum Fisher information before and after the
cosmological decoherence, as well as how the residual quantumness of
the correlations of the final state contribute to it. 

Notice that the periodicity of the saturation of the quantum
Cramer-Rao bound is both in conformal time and field wavenumber. That
is, for a fixed conformal time we can find a specific value of the
field mode that satisfies \eqref{eq:sat}, preserves the total quantum
Fisher information and whose observation allows the highest estimation
precision. These results can be straightforwardly extended to the
multi-parameter case and model-specific conformal factors. From a more
theoretical point of view, since bosonic and fermionic cosmological
particle creation show qualitatively different entanglement behaviors,
it could be interesting to study the same problem for fermionic
fields.

\begin{acknowledgments}
YN was supported in part by JSPS KAKENHI Grant Number 16H01094. MR
  gratefully acknowledges support from the Ministry of Education,
  Culture, Sports, Science and Technology (MEXT) of Japan.
\end{acknowledgments}


\end{document}